\documentstyle[twoside,epsfig]{article}

\catcode`\@=11
\long\def\@makefntext#1{
\protect\noindent \hbox to 3.2pt {\hskip-.9pt  
$^{{\eightrm\@thefnmark}}$\hfil}#1\hfill}               

\def\@makefnmark{\hbox to 0pt{$^{\@thefnmark}$\hss}}    
        
\def\ps@myheadings{\let\@mkboth\@gobbletwo
\def\@oddhead{\hbox{}
\rightmark\hfil\eightrm\thepage}   
\def\@oddfoot{}\def\@evenhead{\eightrm\thepage\hfil
\leftmark\hbox{}}\def\@evenfoot{}
\def\sectionmark##1{}\def\subsectionmark##1{}}

\oddsidemargin=\evensidemargin
\newcounter{sectionc}\newcounter{subsectionc}\newcounter{subsubsectionc}
\renewcommand{\section}[1] {\vspace{12pt}\addtocounter{sectionc}{1} 
\setcounter{subsectionc}{0}\setcounter{subsubsectionc}{0}\noindent 
        {\tenbf\thesectionc. #1}\par\vspace{5pt}}
\renewcommand{\subsection}[1] {\vspace{12pt}\addtocounter{subsectionc}{1} 
        \setcounter{subsubsectionc}{0}\noindent 
        {\bf\thesectionc.\thesubsectionc. {\kern1pt \bfit #1}}\par\vspace{5pt}}
\renewcommand{\subsubsection}[1] {\vspace{12pt}\addtocounter{subsubsectionc}{1}
        \noindent{\tenrm\thesectionc.\thesubsectionc.\thesubsubsectionc.
        {\kern1pt \tenit #1}}\par\vspace{5pt}}
\newcommand{\nonumsection}[1] {\vspace{12pt}\noindent{\tenbf #1}
        \par\vspace{5pt}}

\newcounter{appendixc}
\newcounter{subappendixc}[appendixc]
\newcounter{subsubappendixc}[subappendixc]
\renewcommand{\thesubappendixc}{\Alph{appendixc}.\arabic{subappendixc}}
\renewcommand{\thesubsubappendixc}
        {\Alph{appendixc}.\arabic{subappendixc}.\arabic{subsubappendixc}}

\renewcommand{\appendix}[1] {\vspace{12pt}
        \refstepcounter{appendixc}
        \setcounter{figure}{0}
        \setcounter{table}{0}
        \setcounter{lemma}{0}
        \setcounter{theorem}{0}
        \setcounter{corollary}{0}
        \setcounter{definition}{0}
        \setcounter{equation}{0}
        \renewcommand{\thefigure}{\Alph{appendixc}.\arabic{figure}}
        \renewcommand{\thetable}{\Alph{appendixc}.\arabic{table}}
        \renewcommand{\theappendixc}{\Alph{appendixc}}
        \renewcommand{\thelemma}{\Alph{appendixc}.\arabic{lemma}}
        \renewcommand{\thetheorem}{\Alph{appendixc}.\arabic{theorem}}
        \renewcommand{\thedefinition}{\Alph{appendixc}.\arabic{definition}}
        \renewcommand{\thecorollary}{\Alph{appendixc}.\arabic{corollary}}
        \renewcommand{\theequation}{\Alph{appendixc}.\arabic{equation}}
        \noindent{\tenbf Appendix \theappendixc #1}\par\vspace{5pt}}
\newcommand{\subappendix}[1] {\vspace{12pt}
        \refstepcounter{subappendixc}
        \noindent{\bf Appendix \thesubappendixc. {\kern1pt \bfit #1}}
        \par\vspace{5pt}}
\newcommand{\subsubappendix}[1] {\vspace{12pt}
        \refstepcounter{subsubappendixc}
        \noindent{\rm Appendix \thesubsubappendixc. {\kern1pt \tenit #1}}
        \par\vspace{5pt}}

\newcommand{\textlineskip}{\baselineskip=13pt}
\newcommand{\smalllineskip}{\baselineskip=10pt}

\def\eightcirc{
\begin{picture}(0,0)
\put(4.4,1.8){\circle{6.5}}
\end{picture}}
\def\eightcopyright{\eightcirc\kern2.7pt\hbox{\eightrm c}}

\def\abstracts#1#2#3{{
        \centering{\begin{minipage}{4.5in}\baselineskip=10pt\footnotesize
        \parindent=0pt #1\par
        \parindent=15pt #2\par
        \parindent=15pt #3\par
        \end{minipage}}\par}} 

\newcommand{\bibit}{\nineit}
\newcommand{\bibbf}{\ninebf}
\renewenvironment{thebibliography}[1]
        {\frenchspacing
         \ninerm\baselineskip=11pt
         \begin{list}{\arabic{enumi}.}
        {\usecounter{enumi}\setlength{\parsep}{0pt}     
         \setlength{\leftmargin 17pt}{\rightmargin 0pt}   
         \setlength{\itemsep}{0pt} \settowidth
        {\labelwidth}{#1.}\sloppy}}{\end{list}}

\newcounter{itemlistc}
\newcounter{romanlistc}
\newcounter{alphlistc}
\newcounter{arabiclistc}

\newcommand{\fcaption}[1]{
        \refstepcounter{figure}
        \setbox\@tempboxa = \hbox{\footnotesize Fig.~\thefigure. #1}
        \ifdim \wd\@tempboxa > 5in
           {\begin{center}
        \parbox{5in}{\footnotesize\smalllineskip Fig.~\thefigure. #1}
            \end{center}}
        \else
             {\begin{center}
             {\footnotesize Fig.~\thefigure. #1}
              \end{center}}
        \fi}

\newcommand{\tcaption}[1]{
        \refstepcounter{table}
        \setbox\@tempboxa = \hbox{\footnotesize Table~\thetable. #1}
        \ifdim \wd\@tempboxa > 5in
           {\begin{center}
         \parbox{5in}{\footnotesize\smalllineskip Table~\thetable. #1}
            \end{center}}
        \else
             {\begin{center}
             {\footnotesize Table~\thetable. #1}
              \end{center}}
        \fi}

\def\@citex[#1]#2{\if@filesw\immediate\write\@auxout
        {\string\citation{#2}}\fi
\def\@citea{}\@cite{\@for\@citeb:=#2\do
        {\@citea\def\@citea{,}\@ifundefined
        {b@\@citeb}{{\bf ?}\@warning
        {Citation `\@citeb' on page \thepage \space undefined}}
        {\csname b@\@citeb\endcsname}}}{#1}}

\newif\if@cghi
\def\cite{\@cghitrue\@ifnextchar [{\@tempswatrue
        \@citex}{\@tempswafalse\@citex[]}}
\def\citelow{\@cghifalse\@ifnextchar [{\@tempswatrue
        \@citex}{\@tempswafalse\@citex[]}}
\def\@cite#1#2{{$\null^{#1}$\if@tempswa\typeout
        {IJCGA warning: optional citation argument 
        ignored: `#2'} \fi}}

\def\pmb#1{\setbox0=\hbox{#1}
        \kern-.025em\copy0\kern-\wd0
        \kern.05em\copy0\kern-\wd0
        \kern-.025em\raise.0433em\box0}

\def\fnt#1#2{\footnotetext{\kern-.3em
        {$^{\mbox{\scriptsize #1}}$}{#2}}}

\def\fpage#1{\begingroup
\voffset=.3in
\thispagestyle{empty}\begin{table}[b]\centerline{\footnotesize #1}
        \end{table}\endgroup}

\def\runninghead#1#2{\pagestyle{myheadings}
\markboth{{\protect\footnotesize\it{\quad #1}}\hfill}
{\hfill{\protect\footnotesize\it{#2\quad}}}}
\font\tenbf=cmbx10
\font\tenit=cmti10 
\font\tenit=cmti10
\font\bfit=cmbxti10 at 10pt
\font\ninebf=cmbx9
\font\ninerm=cmr9
\font\nineit=cmti9

\font\eightrm=cmr8

\def\lsym{\raise-3pt\hbox{\vbox{\tabskip0pt\offinterlineskip
        \halign{\tabskip0pt plus 1em
        ##\tabskip0pt\cr
        $\,\,<\,\,$\cr
        $\,\,\sim\,\,$\cr}}}}
\def\rsym{\raise-3pt\hbox{\vbox{\tabskip0pt\offinterlineskip
     \halign{\tabskip0pt plus 1em
      ##\tabskip0pt\cr
      $\,\,>\,\,$\cr
      $\,\,\sim\,\,$\cr}}}}
\def\qed{\hbox{${\vcenter{\vbox{                        
        \hrule height 0.4pt\hbox{\vrule width 0.4pt height 6pt
        \kern5pt\vrule width 0.4pt}\hrule height 0.4pt}}}$}}
\def\theequation{\thesection.\arabic{equation}}         

\newcommand{\Beo}{B_{\mathrm{eo}}}
\newcommand{\Boe}{B_{\mathrm{oe}}}
\newcommand{\Deo}{D_{\mathrm{eo}}}
\newcommand{\Doe}{D_{\mathrm{oe}}}
\newcommand{\Aeo}{A_{\mathrm{eo}}}
\newcommand{\Aoe}{A_{\mathrm{oe}}}
\newcommand{\Ke}{K_{\mathrm{e}}}
\newcommand{\Ko}{K_{\mathrm{o}}}
\newcommand{\trasp}{{\mathrm{T}}}
\newcommand{\tr}{{\mathrm{tr}}}

\begin{document}

\runninghead{J.M. Carmona, M. D'Elia, A. Di Giacomo \& B. Lucini}
{Implementacion of $C^\star$ boundary conditions in the Hybrid Monte 
Carlo algorithm}

\normalsize\textlineskip
\thispagestyle{empty}
\setcounter{page}{1}


\fpage{1}
\centerline{\bf IMPLEMENTATION OF $C^\star$ BOUNDARY CONDITIONS IN THE}
\vspace*{0.035truein}
\centerline{\bf HYBRID MONTE CARLO ALGORITHM} 
\vspace*{0.37truein}
\centerline{\footnotesize JOS\'E MANUEL CARMONA, MASSIMO D'ELIA, ADRIANO 
DI GIACOMO}
\vspace*{0.015truein}
\centerline{\footnotesize\it Dipartimento di Fisica dell'Universit\`a
and INFN, I-56127 Pisa, Italy}
\vspace*{10pt} 
\centerline{\normalsize and}
\vspace*{10pt}
\centerline{\footnotesize BIAGIO LUCINI}
\vspace*{0.015truein}                   
\centerline{\footnotesize\it Theoretical Physics, University of Oxford,}
\baselineskip=10pt
\centerline{\footnotesize\it 1 Keble Road, Oxford, OX1 3NP, UK}  

\vspace*{0.225truein}

\vspace*{0.21truein}
\abstracts{In the study of QCD dynamics, $C^\star$ boundary conditions are 
physically relevant in certain cases. In this paper we study the implementation
of these boundary conditions in the lattice formulation of full QCD with
staggered fermions. In particular, we show that the usual even-odd 
partition trick to avoid the redoubling of the fermion matrix is still 
valid in this case. We give an explicit implementation of these boundary 
conditions for the Hybrid Monte Carlo algorithm.
}{}{}



\vspace*{1pt}\textlineskip      
\section{Introduction}          
\setcounter{section}{1}
\vspace*{-0.5pt}
\noindent
Lattice QCD simulations are usually performed with periodic boundary 
conditions (BC). However, other type of BC may be important in certain cases.
For example, a comparison between systems with different BC can be used
to understand finite size-effects in lattice QCD.
Some years ago\cite{polley-kronfeld} $C$-periodic BC were studied
as an alternative to periodic conditions. Then they were considered
with the general idea of studying the spontaneous symmetry breaking aspects 
of the QCD dynamics in a simple way.\cite{wiese} 
In that work an analysis was done in the
continuum, and it was shown that in pure gauge theory these BC break the
$Z(3)$ symmetry explicitely, which has important consequences for the
high-temperature deconfinement phase transition. These conditions are also
useful in numerical lattice simulations of this transition. When quarks
are present, $C$-periodic BC break both chiral and flavour symmetries.

These boundary conditions are also especially important
when topological properties are relevant in the system under
consideration. This is the case of the lattice studies of confinement through
monopole condensation. Recently the role of monopoles in
connection with colour confinement has been evidentiated
in SU(2) and SU(3) gluodynamics,\cite{monopole} for which a disorder
parameter based on the magnetic U(1) symmetry has been constructed and
studied by Monte Carlo techniques. The disorder parameter is
the vacuum expectation value ({\em vev}) of a disorder operator,
which is an operator that creates a magnetic monopole in the
gauge configuration. The definition
of this disorder operator requires $C$-periodic BC in the time direction.
For the pure gauge case, this means that the links at time $t+N_t$, where
$N_t$ is the temporal extension of the lattice, are the complex conjugate
of the links at time $t$. The effect on the
simulation algorithm is a simple redefinition of the staples containing links
that pierce the temporal boundary.
The natural extension of the procedure used for the pure gauge case to full
QCD requires the implementation of $C$-periodic BC in the presence of fermions.
In particular, we will be concerned with the case of staggered fermions.
$C$-periodic BC modify the fermionic matrix, and many proofs of properties
used for the setup of standard simulation algorithms no longer hold.

$C$-periodic BC in the continuum are 
defined by the action of the charge conjugation operator $C$ on the 
fields.\cite{wiese} However, lattice fermions are different from fermions in
the continuum. In particular, $C$ is not a symmetry of the lattice action
with staggered fermions. It also breaks translation invariance for finite 
lattice spacing. However, there is a discrete symmetry of the staggered 
fermion action,
\begin{equation}
S_f=\sum_{i,\mu} \left[\frac{1}{2}\,\eta_{i,\mu} 
\left(\bar\psi_i U_{i,\mu} \psi_{i+\mu} - \bar\psi_{i+\mu}
U^\dag_{i,\mu} \psi_i\right) + m \bar\psi_i\psi_i\right]
\label{staggaction}
\end{equation}
(here $i$ indicates the lattice point, $U_{i,\mu}$ is the SU(3) matrix
associated with the link leaving the $i$-th lattice point in the $\mu$
direction, $\psi_i$ is the staggered fermion field at the point $i$, and 
$\eta_{i,\mu}$ the usual staggered fermion phase),  
which corresponds to charge conjugation in
the gluon sector but which in the continuum limit also contains a flavour
transformation.\cite{wiese,golterman} We will call $C^\star$ 
this symmetry of the lattice action:
\begin{equation}
^{C^\star} U_{i,\mu}=U^*_{i,\mu}, \quad 
^{C^\star}\psi_i=\epsilon_i\bar\psi^\trasp_i,
\quad ^{C^\star}\bar\psi_i=-\psi^\trasp_i\epsilon_i,
\label{symmetry}
\end{equation}
where $\epsilon_i=(-1)^{x_i+y_i+z_i+t_i}$, $(x_i,y_i,z_i,t_i)$ being the 
lattice coordinates of point $i$, the ``T'' represents the traspose operation,
and we will use $\psi^* \equiv \bar\psi^\trasp$.
Translation invariance implies that a BC must correspond to a symmetry of the
action. $C^\star$-BC are defined as the boundary conditions corresponding to
the symmetry~(\ref{symmetry}):
\begin{equation}
\Phi_{i+N} = \,^{C^\star}\Phi_i,
\label{BC}
\end{equation}
where $\Phi$ is a field, $U_\mu$ or $\psi$, and $N$ is the number of lattice 
points in the direction in which we use this boundary condition. 

In the chiral limit the lattice action has a $U(1)_{\mathrm{E}}\otimes
U(1)_{\mathrm{O}}$ chiral symmetry of independent rotations on 
$(x+y+z+t)$-even and -odd lattice points. $C^\star$-BC break explicitely
this symmetry to $U(1)_{\mathrm{E=O^*}}$. Baryon number 
$U(1)_{\mathrm{E=O}}$ is also broken explicitely down to
$Z(2)_{\mathrm{E=O}}$.\cite{wiese} 

In this paper we will have in mind the physical problem 
mentioned above: computation of the {\em vev} of a monopole creation operator 
in lattice QCD.
This means that we will consider imposing $C^\star$-BC in 
the time direction, and periodic BC in the spatial directions.
However this is just to fix the notation in what follows;
the $C^\star$ conditions could in fact be assumed in any direction. 
The purpose of the 
paper is to show the theoretical framework to be used in a lattice
simulation with these BC (section 2) and how usual algorithms\cite{gottlieb}
need to be modified (section 3). Our conclusions are summarized in section 4.

\setcounter{equation}{0}
\setcounter{section}{2}
\section{Mathematical description}
\noindent 
Let us consider the partition function of lattice QCD with staggered fermions 
\begin{equation}  
Z = \int ({\cal D} U) ({\cal D} \psi {\cal D} \bar \psi) \,e^{-S_g -S_f}\ ,
\end{equation}
where $S_g(U)$ is the Wilson action for the pure gauge sector, and $S_f$
is given by Eq.~(\ref{staggaction}). The fermionic variables can be 
integrated out to give
\begin{equation}
Z = \int ({\cal D} U)\, e^{-S_g(U)} \, \det M(U) \ ,
\label{dynfermions}
\end{equation}
where\footnote{We have absorbed the staggered phases by a redefinition of the
link matrices $U$.}
\begin{equation}
M(U)_{i,j}=m \delta_{i,j} + \sum_{\mu} \frac{1}{2}\,
(U_{i,\mu}\delta_{i,j-\mu}-U^\dag_{i-\mu,\mu}\delta_{i,j+\mu})
\label{fmatrix}
\end{equation}
is the fermionic matrix, and periodic BC are assumed. 
Using this matrix notation, we can write
\begin{equation}
S_f=\bar\psi M \psi=\frac{1}{2} \left[\left(\bar\psi M\psi\right)+
\left(\bar\psi M \psi\right)^\trasp\right],
\label{faction}
\end{equation}
since $S_f$ is a number. We can use the new variable $\Psi$ defined
as the column vector formed by $\psi$ and $\psi^*$, so that the fermionic
integral is
\begin{equation}
\label{compzeta}
\int ({\cal D} \psi) ({\cal D} \bar \psi) \,e^{-S_f}=
\int ({\cal D} \Psi) \, 
e^{-\frac{1}{2}\Psi^\trasp A \Psi} = \mathrm{Pf}(A) \ ,
\label{fintegral}
\end{equation}
\begin{equation}
\Psi^\trasp A\Psi \equiv 
\left(\begin{array}{cc} \psi^\trasp & \bar\psi\end{array}\right)
A \left(\begin{array}{l} \psi \\ \psi^*\end{array}\right) \ ,
\label{basechoice}
\end{equation}
with 
\begin{equation}
A=\left(\begin{array}{cc} 0 & -M^\trasp \\ M & 0 \end{array}
\right) \ ,
\label{matrixA}
\end{equation}
and we have introduced the Pfaffian of the matrix $A$, $\mathrm{Pf}(A)$.
It is well known that\cite{zj}
\begin{equation}
\mathrm{Pf}^2(A) = \det A \ .
\label{pf}
\end{equation}

Now let us consider $C^\star$-BC in the time direction. Then, following
Eqs.~(\ref{symmetry}) and~(\ref{BC}), 
$\psi_{N_t}=\epsilon_0\psi^*_0$, and 
$\bar\psi_{N_t}=-\epsilon_0\psi^\trasp_0$, where we have written
in the subscript the temporal coordinate and omitted the spatial coordinates. 
Now Eq.~(\ref{faction}) gives, for the terms connecting the slices $N_t-1$
and $N_t$,
\begin{eqnarray}
&&\frac{1}{2} \left(\bar\psi_{N_t-1}U_{N_t-1,t}\psi^*_0 +
\psi^\trasp_0 U^\dag_{N_t-1,t}\psi_{N_t-1}\right)\epsilon_0 \nonumber \\
&-&\frac{1}{2} \left(\bar\psi_0 U^\trasp_{N_t-1,t}\psi^*_{N_t-1} +
\psi^\trasp_{N_t-1} U^*_{N_t-1,t}\psi_0\right)\epsilon_0 \, .
\label{borderterm}
\end{eqnarray}
In this way the matrix $A$ of Eq.~(\ref{matrixA}) is substituted by
\begin{equation}
A=\left(\begin{array}{cc} B & -\tilde M^\trasp \\ 
\tilde M & -B^* \end{array}\right),
\label{newmatrixA}
\end{equation}
where $B$ satisfies the properties:
\begin{equation}
B^\dag=-B^* \quad \quad B=-B^\trasp
\label{Bprops}
\end{equation}
and $\tilde M$ is the fermionic matrix Eq.~(\ref{fmatrix}) apart from the
terms connecting the slices $N_t-1$ and $N_t$, which have gone to the matrices
$B$ and $-B^*$. 

Eq.~(\ref{fintegral}) is still valid and we are interested in
calculating $\mathrm{Pf}(A)$. 


\subsection{Pseudofermionic variables}
\noindent
The usual approach\cite{polonyi} to the simulation of theories 
with dynamical fermions
is to rewrite the determinant of the fermionic matrix in 
Eq.~(\ref{dynfermions}) using that
\begin{equation}
\det (M^\dag M) \propto \int {\cal D}\phi^\dag {\cal D}\phi
\, \exp\,[-\phi^\dag(M^\dag M)^{-1}\phi] \ ,
\end{equation}
where $\phi$ is a complex bosonic field with the same quantum numbers
as the Grassmann field. 
One introduces the matrix $(M^\dagger M)^{-1}$, instead of 
$M^{-1}$, so that the pseudofermionic fields can be generated
using a simple heatbath method.
Since $\det M$ is a
real number, $\det (M^\dag M)=(\det M)^2$. 
So, actually, this corresponds to a double number of flavours with 
respect to the original theory. However, 
the matrix $M^\dag M$ has two important 
properties: it has no matrix elements connecting even and odd lattice
sites, and the determinants of its submatrices on the even and odd sites
are equal.\cite{martin} Therefore one can avoid the redoubling of flavours
by defining the pseudofermionic field only on even lattice sites.

A remarkable difference between the partition function with periodic BC and the
partition function with $C^{\star}$-BC is that in the latter case the
determinant of $M$, Eq.~(\ref{dynfermions}), has to be replaced by
the Pfaffian of $A$, i.e. $\pm \sqrt{\det A}$. Because of the square root,
the usual trick of introducing pseudofermionic fields and rewriting this
factor as the integral over
these fields can only be applied if the number of continuum fermion flavours is
such that the square root cancels. Moreover, in order to have a
positive-definite integration measure, we need that the sign in front of
this factor be $+$. Both conditions are satisfied if the number of continuum
flavours is a multiple of eight. This generates a further unavoidable
redoubling. Until Sect. 3.2, we will be concerned with the numerical
simulation of $\det A$, which is equivalent to simulating a double number of
fermion flavours in the continuum limit (eight instead of four). In Sect.
3.2 we will discuss how to deal with the usual case of four staggered fermion
flavours.

Since the case of a system with eight fermion flavours in the continuum
limit and $C^\star$-BC and the case of a system with four fermion flavours
in the continuum limit and periodic BC are similar, we would like to follow
in the former case the standard procedure to obtain the determinant of
the matrix~(\ref{newmatrixA}). First, in the Appendix A it is shown that
$\det A$ is a real number. So we can also in this case use the
matrix $A^\dag A$ to introduce the pseudofermionic field, which will have 
now twice the number of components as in the usual case.
Second, we will now see that the matrix $A^\dag A$ does not connect 
even and odd lattice sites.

Using the form of the fermionic matrix~(\ref{fmatrix}) and the definitions
of $B$ and $\tilde M$ given by Eqs.~(\ref{borderterm}) and~(\ref{newmatrixA}),
we can split the blocks of the matrix $A$ in even and odd lattice sites:
\begin{equation}
A=\left(\begin{array}{cccc} 0 & 
\frac{1}{2}\Beo & -m & -\frac{1}{2} \Doe^\trasp \\
\frac{1}{2}\Boe & 0 & -\frac{1}{2}\Deo^\trasp & -m \\
m & \frac{1}{2}\Deo & 0 & -\frac{1}{2}\Beo^* \\
\frac{1}{2}\Doe & m & -\frac{1}{2}\Boe^* & 0 \end{array} \right).
\label{blocks}
\end{equation}
From Eqs.~(\ref{fmatrix}) 
and~(\ref{Bprops}) it is easy to see that
\begin{eqnarray} 
\Doe^\dag=-\Deo && \Doe^\trasp=-\Deo^* \label{Doeprop} \\
\Boe^\dag=-\Beo^* && \Boe^\trasp=-\Beo  \label{Boeprop} \ .
\end{eqnarray}

In the expression~(\ref{blocks}), the new blocks divide the matrix in 
rows and columns identified by the pairs $[0,\mathrm{e}]$, $[0,\mathrm{o}]$, 
$[1,\mathrm{e}]$ and $[1,\mathrm{o}]$, where the number indicates the blocks
defined in Eq.~(\ref{newmatrixA}) and the letter the even/odd subblock.
Reorganizing the rows and columns to $[0,\mathrm{e}]$, $[1,\mathrm{e}]$, 
$[0,\mathrm{o}]$ and $[1,\mathrm{o}]$ (the determinant does not change), 
we rewrite the matrix A in the following form:
\begin{equation}
A=\left(\begin{array}{cc} \mu & \frac{1}{2}\Aeo \\
\frac{1}{2}\Aoe & \mu \end{array}\right),
\label{Areorg}
\end{equation}
where
\begin{equation}
\mu=\left(\begin{array}{cc} 0 & -m \\
m & 0 \end{array}\right),
\end{equation}
\begin{equation}
\Aeo=\left(\begin{array}{cc} \Beo & \Deo^* \\
\Deo & -\Beo^* \end{array}\right),
\label{Aeosubblocks}
\end{equation}
and $\Aoe$ has the same form as $\Aeo$ with the exchange 
$\mathrm{e}\leftrightarrow\mathrm{o}$. 
Eqs.~(\ref{Doeprop}) and~(\ref{Boeprop}) give
\begin{equation}
\Aoe^\dag=-\Aeo^* \ .
\label{Aoeprop}
\end{equation}

From Eqs.~(\ref{fmatrix}) and~(\ref{borderterm}) we obtain the explicit
form of the matrices $\Deo$ and $\Beo$:
\begin{equation}
(\Deo)_{j,k}=\sum_\mu (U_{j,\mu}\delta_{k,j+\mu}-
U^\dag_{j-\mu,\mu}\delta_{k,j-\mu}) \ ,
\label{Deoexplicit}
\end{equation}
where $j$ is an even site and the link between $j$ and $k$ does not
connect the $t = N_t -1$ and $t = 0$ time slices 
(otherwise $(\Deo)_{j,k} = 0$), and
\begin{equation}
\left.
\begin{array}{lr}
(\Beo)_{j,k}=U^*_{j,t} & j=N_t-1, k=0 \\
(\Beo)_{j,k}=U^\dag_{k,t} & j=0, k=N_t-1 \\
(\Beo)_{j,k}=0 & \mbox{in every other case}
\end{array}
\right\}
\end{equation}
where $j=N_t-1$ or $j=0$ means that the temporal coordinate of site $j$
is $N_t-1$ or zero, respectively. When $j$ is an odd site, $(\Doe)_{j,k}$
has the same expression as Eq.~(\ref{Deoexplicit}), and
\begin{equation}
\left.
\begin{array}{lr}
(\Boe)_{j,k}=-U^*_{j,t} & j=N_t-1, k=0 \\
(\Boe)_{j,k}=-U^\dag_{k,t} & j=0, k=N_t-1 \\
(\Boe)_{j,k}=0 & \mbox{in every other case}
\end{array}
\right\}
\label{Boeexplicit}
\end{equation}
These expressions completely determine every element of the matrices
$\Aeo$ and $\Aoe$.

We now compute
\begin{equation}
A^\dag A=\left(\begin{array}{cc} -\mu^2+\frac{1}{4}\Aoe^\dag\Aoe &
-\frac{1}{2}(\mu\Aeo-\Aoe^\dag\mu) \\
\frac{1}{2}(\Aeo^\dag\mu-\mu\Aoe) &
-\mu^2+\frac{1}{4}\Aeo^\dag\Aeo \end{array}\right).
\label{AcroceA1}
\end{equation}
But using the properties~(\ref{Doeprop}) and~(\ref{Boeprop}) it is direct
to see that the blocks out of the diagonal in Eq.~(\ref{AcroceA1}) are zero.
Then, using Eq.~(\ref{Aoeprop}),
\begin{equation}
A^\dag A=\left(\begin{array}{cc} -\mu^2-\frac{1}{4}\Aeo^*\Aoe & 0 \\
0 & -\mu^2-\frac{1}{4}\Aoe^*\Aeo \end{array}\right)
\label{AcroceA2}
\end{equation}
and, as a result, we see that the matrix $A^\dag A$ connects only lattice 
points of the same parity.

In order to use the same trick to avoid the flavour doubling produced by
the introduction of $A^\dag A$, that is, to define the pseudofermionic field on
even sites only, we need to show that, also in this case, the determinant
of the even and odd parts in the matrix~(\ref{AcroceA2}) are equal. This is
done in the following subsection.


\subsection{Reducing the number of flavours: even-odd partitioning}
\noindent
Let us write $K\equiv A^\dag A$. Then we have
\begin{equation}
K=\left(\begin{array}{cc} m^2 I_2-\frac{1}{4}\Aeo^*\Aoe & 0 \\
0 & m^2 I_2-\frac{1}{4}\Aoe^*\Aeo \end{array}\right)\equiv
\left(\begin{array}{cc} \Ke & 0 \\ 0 & \Ko \end{array}\right), 
\end{equation}
where $I_2$ is the $2\times 2$ identity matrix. We need to show that
$\det\Ke=\det\Ko$. The strategy
will be to show that both $\det\Ke$ and $\det\Ko$ are equal
to $\det A$.
To this aim, let us consider the matrix $A$ written in the e--o 
partitioned form, Eq. (\ref{Areorg}),
\begin{equation}
A=\left(\begin{array}{cc} \mu & \frac{1}{2}\Aeo \\
\frac{1}{2}\Aoe & \mu \end{array}\right),
\label{Areorg2}
\end{equation}
where
\begin{equation}
\mu=\left(\begin{array}{cc} 0 & -m \\
m & 0 \end{array}\right),
\end{equation}
\begin{equation}
\Aeo=\left(\begin{array}{cc} \Beo & \Deo^* \\
\Deo & -\Beo^* \end{array}\right),
\end{equation}
and
\begin{equation}
\Aoe=\left(\begin{array}{cc} \Boe & \Doe^* \\
\Doe & -\Boe^* \end{array}\right).
\end{equation}

We will make use of a general property of the determinant of a square
$2N \times 2N$ block matrix,
\begin{equation}
\det \left(\begin{array}{cc} X & Y\\ 
W & Z \end{array}\right) = \det (X) \det(Z - W X^{-1} Y)\;,
\label{blockexp}
\end{equation}
where $X,Y,W,Z$ are $N \times N$ square matrices and $X$ is invertible.
Applying this equality to the computation of $\det A$, we obtain
\begin{equation}
\det A  = \det(\mu) 
             \det \left(\mu - \frac{1}{4} \Aoe \mu^{-1} \Aeo \right) =  
         \det \left(\mu^2 - \frac{1}{4} \mu \Aoe \mu^{-1} \Aeo \right).
\label{detaoe}
\end{equation}
Using the explicit form of $\mu$ and $\Aoe$ it is easily shown that
\begin{equation}
\mu \Aoe = -  \Aoe^* \mu\;.
\label{permu}
\end{equation}
Inserting this equality in Eq. (\ref{detaoe}), it follows that
\begin{equation}
\det A =  \det \left(\mu^2 + \frac{1}{4} \Aoe^* \Aeo \right)
       =  \det \Ko \ ,
\end{equation}
where the last equality is implied by the fact that $N$ is an even number.

We can compute again $\det A$ using the same procedure after exchanging
even and odd variables in $A$. Using a property analogous to that
in Eq. (\ref{permu}), namely $\mu \Aeo = - \Aeo^* \mu$,
it is then easily shown that
\begin{equation}
\det A =  \det \left(\mu^2 + \frac{1}{4} \Aeo^* \Aoe \right)
       =  \det \Ke \ . 
\end{equation}

This is a proof that $\det \Ke = \det \Ko$. 

We have
shown that $\det\Ke \det\Ko = (\det A)^2$. On the other
hand it is also true that  
$\det\Ke \det\Ko = \det (A^\dag A) =  \det A^\dag \det A$,
therefore we have also obtained a proof 
that $\det A = \det A^\dag$ alternative
to that given in Appendix A. Moreover, we have that
\begin{equation}
\det A= \det \Ke = \det \left(m^2 I_2 + \frac{1}{4} \Aoe^\dag \Aoe \right)
> 0 ,
\label{detApositive}
\end{equation}
since both $m^2 I_2$ and $\Aoe^\dag \Aoe$ are positive definite matrices.
Therefore $\det A$ is a positive number.

We notice that this method can be easily applied also to the standard
case, thus providing a proof alternative to those presented  in Ref.~8.
This is shown in detail in Appendix B.

\setcounter{equation}{0}
\setcounter{section}{3}
\section{Hybrid Monte Carlo Implementation}
\noindent
We will show now how the standard Hybrid Monte Carlo (HMC) 
algorithm\cite{gottlieb} needs to be modified to incorporate $C^\star$-BC.
In this algorithm one introduces fictitious momenta, conjugate variables
of the links, as dynamical variables, and makes \mbox{fields} evolve
with a mixed dynamics, in which deterministic
and stochastic steps are alternated in a prescribed way. In the deterministic
part of the algorithm, the system follows the equations of motion derived
from the Hamiltonian of the (4+1)-dimensional system. These equations give
the evolution of the fields in the fictitious time $\tau$. The 
equation of motion for the matrix $U_{j,\mu}$ is
\begin{equation}
\dot{U}_{j,\mu}=i H_{j,\mu}U_{j,\mu} \ ,
\label{Uevol}
\end{equation}
where the conjugate momentum $H_{j,\mu}$ is a traceless Hermitian matrix, 
and $\dot{U}$ is the
derivative of $U$ with respect to $\tau$. The equations of motion for the
momenta $H$ are obtained by imposing that the Hamiltonian be constant.
The integration of these equations of motion is carried out numerically 
after discretization of $\tau$. Usually the temporal step is of order
$10^{-2}$--$10^{-3}$ and the configuration space is sampled with a total
length of the trajectory of order 1. The stochastic part of the algorithm
consists in the generation of new momenta and new pseudofermionic variables
according to their probability distributions at the beginning of each
trajectory. Moreover, at the end of each trajectory a Metropolis
accept-reject step is performed, which makes the algorithm exact.

\subsection{HMC algorithm with $C^\star$ boundary conditions}
\noindent
Once introduced the pseudofermionic fields, defined only on even sites,
and the auxiliary momenta fields, the partition function of the system is
\begin{equation}
Z = \int ({\cal D}U {\cal D}\phi^\dag {\cal D}\phi {\cal D}H)
\, e^{-\cal{H}}  \ ,
\end{equation} 
with
\begin{equation}
{\cal H}=\frac{1}{2}\sum_{j,\mu} \tr H^2_{j,\mu} + S_g +
\phi^\dag (A^\dag A)^{-1} \phi \ .
\label{Hamiltonian}
\end{equation}
To obtain an equation of motion for $H$ we require that ${\cal H}$ be a
constant of motion, that is, $\dot{\cal H}=0$. The tricky part in this
differentiation is in the fermionic term. The 
derivative of ${\cal H}_f=\phi^\dag (A^\dag A)^{-1} \phi$ is
\begin{eqnarray}
\dot{\cal H}_f&=&-\sum_{j,\mu}\phi^\dag (A^\dag A)^{-1} \,i \left[
\frac{\partial A^\dag}{\partial U_{j,\mu}} H_{j,\mu} U_{j,\mu} A +
\frac{\partial A^\dag}{\partial U^\trasp_{j,\mu}} U^\trasp_{j,\mu} 
H^\trasp_{j,\mu} A + \right. \nonumber \\
&& A^\dag \frac{\partial A}{\partial U_{j,\mu}} H_{j,\mu} U_{j,\mu} + 
A^\dag \frac{\partial A}{\partial U^\trasp_{j,\mu}} 
U^\trasp_{j,\mu} H^\trasp_{j,\mu} -
H^\trasp_{j,\mu} U^*_{j,\mu} \frac{\partial A^\dag}{\partial U^*_{j,\mu}} A -
\nonumber \\
&& \left.
U^\dag_{j,\mu} H_{j,\mu} \frac{\partial A^\dag}{\partial U^\dag_{j,\mu}} A -
A^\dag H^\trasp_{j,\mu} U^*_{j,\mu}\frac{\partial A}{\partial U^*_{j,\mu}} -
A^\dag U^\dag_{j,\mu} H_{j,\mu} \frac{\partial A}{\partial U^\dag_{j,\mu}}
\right] \nonumber \\
&&\times (A^\dag A)^{-1} \phi \ ,
\end{eqnarray} 
where we have made use of Eq.~(\ref{Uevol}) and the fact that $H$ is
Hermitian. We have also used that $A$ is linear in $U$, $U^\trasp$, $U^*$
and $U^\dag$, and then the partial derivatives in the previous
expression commute with $H$ and $U$.

It is convenient to introduce the operator
\begin{equation}
P_{ij}=X_i X^\dag_j \ ,
\end{equation}
where
\begin{equation}
X=(A^\dag A)^{-1}\phi \ .
\end{equation}
It is easily seen that $P$ is Hermitian:
\begin{equation}
(P^\dag)_{ij}=(P_{ji})^\dag=X_i X^\dag_j=P_{ij} \ ,
\end{equation}
and, being $\phi$ defined only on even sites, $P_{ij}$ is taken to be
zero unless $i$ and $j$ are both even sites. On the other hand,
we have that
\begin{eqnarray} 
\left(\frac{\partial A^\dag}{\partial U}\right)^\dag = 
\frac{\partial A}{\partial U^\dag} && 
\left(\frac{\partial A}{\partial U}\right)^\dag = 
\frac{\partial A^\dag}{\partial U^\dag} \nonumber \\
\left(\frac{\partial A^\dag}{\partial U^\trasp}\right)^\dag = 
\frac{\partial A}{\partial U^*} && 
\left(\frac{\partial A}{\partial U^\trasp}\right)^\dag = 
\frac{\partial A^\dag}{\partial U^*}
\end{eqnarray}
and then we can write
\begin{eqnarray}
\dot{\cal H}_f&=&-\sum_{j,\mu}
\tr \left[i H_{j,\mu}\left(U_{j,\mu} A P 
\frac{\partial A^\dag}{\partial U_{j,\mu}}
+U_{j,\mu} P A^\dag \frac{\partial A}{\partial U_{j,\mu}}\right) +
\nonumber \right.\\
&& \left. i H^\trasp_{j,\mu}\left(A P 
\frac{\partial A^\dag}{\partial U^\trasp_{j,\mu}} U^\trasp_{j,\mu}
+P A^\dag \frac{\partial A}{\partial U^\trasp_{j,\mu}} U^\trasp_{j,\mu} 
\right) + \mathrm{H.c.}\right] ,
\label{derHf}
\end{eqnarray}
where H.c. means the Hermitian conjugate and we have used the cyclic
property of the trace operation.

The trace in Eq.~(\ref{derHf}) is taken over both color and site indices.
Taking this into account and using the cyclic property again, we have
\begin{eqnarray}
\dot{\cal H}_f&=&-\sum_{j,\mu}
\tr \left(H_{j,\mu}\left\{i\left[
U_{j,\mu}\left(A P \frac{\partial A^\dag}{\partial U_{j,\mu}}\right)+
U_{j,\mu}\left(P A^\dag \frac{\partial A}{\partial U_{j,\mu}}\right)+
\right.\right.\right.\nonumber \\
&& \left.\left.\left.
U_{j,\mu}\left(A P 
\frac{\partial A^\dag}{\partial U^\trasp_{j,\mu}}\right)^\trasp +
U_{j,\mu}\left
(P A^\dag \frac{\partial A}{\partial U^\trasp_{j,\mu}}\right)^\trasp 
\right] + \mathrm{H.c.}\right\}\right).
\label{derHf2}
\end{eqnarray}

The following step is to calculate the derivatives appearing
in Eq.~(\ref{derHf2}). From Eqs.~(\ref{Areorg}) and~(\ref{Aoeprop}), we
have that 
\begin{equation}
A^\dag = -A^* \ .
\end{equation}
We write again $A$ and $A^\dag$ in terms of subblocks, identifying the
rows and columns with the pairs $[0,\mathrm{e}]$, $[1,\mathrm{e}]$, 
$[0,\mathrm{o}]$ and $[1,\mathrm{o}]$:
\begin{equation}
A=\left(\begin{array}{cccc} 0 & -m &
\frac{1}{2}\Beo & \frac{1}{2} \Deo^* \\
m & 0 & \frac{1}{2}\Deo & -\frac{1}{2}\Beo^* \\
\frac{1}{2}\Boe & \frac{1}{2}\Doe^* & 0 & -m \\
\frac{1}{2}\Doe & -\frac{1}{2}\Boe^* & m & 0 \end{array} \right),
\label{Asubblocks}
\end{equation}
\begin{equation}
A^\dag=\left(\begin{array}{cccc} 0 & m &
-\frac{1}{2}\Beo^* & -\frac{1}{2} \Deo \\
-m & 0 & -\frac{1}{2}\Deo^* & \frac{1}{2}\Beo \\
-\frac{1}{2}\Boe^* & -\frac{1}{2}\Doe & 0 & m \\
-\frac{1}{2}\Doe^* & \frac{1}{2}\Boe & -m & 0 \end{array} \right).
\end{equation}
Once written everything in terms of $B$ and $D$ we can easily see
where $U$ and $U^\trasp$ appear, from 
Eqs.~(\ref{Deoexplicit})--(\ref{Boeexplicit}), and compute the
derivatives. The result is given in Table~\ref{table:derivatives}.

\begin{table}[htbp]
\tcaption{Value of the derivatives appearing in Eq.~(\ref{derHf2}).
$\mu=t$ indicates the time direction, in which we are imposing the
$C^\star$-BC.}
\smallskip
\label{table:derivatives}
\centerline{\footnotesize\smalllineskip
\begin{tabular}{|c|c|c|c|}
\hline  \multicolumn{2}{|c}{} & $j$ even & $j$ odd  \\
\hline \hline
{} & \begin{tabular}{c}$j\neq N_t-1$ \\ or $\mu\neq t$ \end{tabular} &
{\large $\frac{1}{2} \, \delta_{\alpha,[1,{\mathrm{e}}]j} \,
\delta_{\beta,[0,{\mathrm{o}}]j+\mu}$} &
{\large $\frac{1}{2} \, \delta_{\alpha,[1,{\mathrm{o}}]j} \,
\delta_{\beta,[0,{\mathrm{e}}]j+\mu}$} \\
\cline{2-4}
{\raisebox{2.5ex}[0pt]{\large $\left(\frac{\partial A}{\partial U_{j,\mu}}\right)_{\alpha,\beta}$}} 
& \begin{tabular}{c}$j=N_t-1$ \\ and $\mu=t$ \end{tabular} &
{\large $-\frac{1}{2} \, \delta_{\alpha,[1,{\mathrm{e}}]j} \,
\delta_{\beta,[1,{\mathrm{o}}]j+\mu}$} &
{\large $\frac{1}{2} \, \delta_{\alpha,[1,{\mathrm{o}}]j} \,
\delta_{\beta,[1,{\mathrm{e}}]j+\mu}$} \\
\hline \hline
{} & \begin{tabular}{c}$j\neq N_t-1$ \\ or $\mu\neq t$ \end{tabular} &
{\large $-\frac{1}{2} \, \delta_{\alpha,[0,{\mathrm{e}}]j} \,
\delta_{\beta,[1,{\mathrm{o}}]j+\mu}$} &
{\large $-\frac{1}{2} \, \delta_{\alpha,[0,{\mathrm{o}}]j} \,
\delta_{\beta,[1,{\mathrm{e}}]j+\mu}$} \\
\cline{2-4}
{\raisebox{2.5ex}[0pt]{\large $\left(\frac{\partial A^\dag}
{\partial U_{j,\mu}}\right)_{\alpha,\beta}$}} 
& \begin{tabular}{c}$j=N_t-1$ \\ and $\mu=t$ \end{tabular} &
{\large $-\frac{1}{2} \, \delta_{\alpha,[0,{\mathrm{e}}]j} \,
\delta_{\beta,[0,{\mathrm{o}}]j+\mu}$} &
{\large $\frac{1}{2} \, \delta_{\alpha,[0,{\mathrm{o}}]j} \,
\delta_{\beta,[0,{\mathrm{e}}]j+\mu}$} \\
\hline \hline
{} & \begin{tabular}{c}$j\neq N_t-1$ \\ or $\mu\neq t$ \end{tabular} &
{\large $-\frac{1}{2} \, \delta_{\alpha,[0,{\mathrm{o}}]j+\mu} \,
\delta_{\beta,[1,{\mathrm{e}}]j}$} &
{\large $-\frac{1}{2} \, \delta_{\alpha,[0,{\mathrm{e}}]j+\mu} \,
\delta_{\beta,[1,{\mathrm{o}}]j}$} \\
\cline{2-4}
{\raisebox{2.5ex}[0pt]{\large $\left(\frac{\partial A}
{\partial U^\trasp_{j,\mu}}\right)_{\alpha,\beta}$}} 
& \begin{tabular}{c}$j=N_t-1$ \\ and $\mu=t$ \end{tabular} &
{\large $\frac{1}{2} \, \delta_{\alpha,[1,{\mathrm{o}}]j+\mu} \,
\delta_{\beta,[1,{\mathrm{e}}]j}$} &
{\large $-\frac{1}{2} \, \delta_{\alpha,[1,{\mathrm{e}}]j+\mu} \,
\delta_{\beta,[1,{\mathrm{o}}]j}$} \\
\hline \hline
{} & \begin{tabular}{c}$j\neq N_t-1$ \\ or $\mu\neq t$ \end{tabular} &
{\large $\frac{1}{2} \, \delta_{\alpha,[1,{\mathrm{o}}]j+\mu} \,
\delta_{\beta,[0,{\mathrm{e}}]j}$} &
{\large $\frac{1}{2} \, \delta_{\alpha,[1,{\mathrm{e}}]j+\mu} \,
\delta_{\beta,[0,{\mathrm{o}}]j}$} \\
\cline{2-4}
{\raisebox{2.5ex}[0pt]{\large $\left(\frac{\partial A^\dag}
{\partial U^\trasp_{j,\mu}}\right)_{\alpha,\beta}$}} 
& \begin{tabular}{c}$j=N_t-1$ \\ and $\mu=t$ \end{tabular} &
{\large $\frac{1}{2} \, \delta_{\alpha,[0,{\mathrm{o}}]j+\mu} \,
\delta_{\beta,[0,{\mathrm{e}}]j}$} &
{\large $-\frac{1}{2} \, \delta_{\alpha,[0,{\mathrm{e}}]j+\mu} \,
\delta_{\beta,[0,{\mathrm{o}}]j}$} \\
\hline
\end{tabular}}
\end{table}

Now it is easy to calculate the different
terms in Eq.~(\ref{derHf2}). Let us do, as an example, the first one.
The trace over lattice site indices affects only the expressions between
parentheses. From Table~\ref{table:derivatives}, we see that, for $j$ even,
and $j\neq N_t-1$ or $\mu\neq t$, the first one gives
\begin{eqnarray}
\tr \left(A P \frac{\partial A^\dag}{\partial U_{j,\mu}}\right)&=&
A_{k_2,k} X_k X^\dag_{k_1} \left(\frac{\partial A^\dag}
{\partial U_{j,\mu}}\right)_{k_1,k_2}=
-\frac{1}{2} A_{[1,{\mathrm{o}}]j+\mu,k} X_k X^\dag_{[0,{\mathrm{e}}]j}=
\nonumber \\
&& \left(-\frac{1}{4}(\Doe)_{j+\mu,k} X^0_k + 
\frac{1}{4}(\Boe^*)_{j+\mu,k} X^1_k \right) (X^\dag)^0_j =
\nonumber \\
&& -\frac{1}{4}((\Aoe)_{j+\mu,k} X_k)^1 (X^\dag)^0_j = 
-\frac{1}{4} (\Aoe X)^1_{j+\mu} (X^\dag)^0_j \ ,
\end{eqnarray}
where we have used Eqs.~(\ref{Asubblocks}) and~(\ref{Aeosubblocks}),
and called $X^0_k$ the first three components of the vector
$X_k$, and $X^1_k$ the second three components. 
Notice that there is not a mass term because it is diagonal, which would
imply that $k=j+\mu$, that is, $k$ would be odd, and then $X_k=0$.
For $j$ even and $j=N_t-1$, $\mu=t$, an analogous calculation gives
\begin{equation}
\tr \left(A P \frac{\partial A^\dag}{\partial U_{j,\mu}}\right)=
-\frac{1}{4} (\Aoe X)^0_{j+\mu} (X^\dag)^0_j \ .
\end{equation}
This term does not contribute when $j$ is odd, because in this case 
$(X^\dag)^0_j$ is zero ($X$ is defined on even sites only).

Computing every other contribution in Eq.~(\ref{derHf2}), the final result
can be written
\begin{equation}
\dot{\cal H}_f=-\sum_{j,\mu}
\tr \left\{H_{j,\mu}\left(i F_{j,\mu}- i F^\dag_{j,\mu}\right)\right\} \ ,
\end{equation}
where $F_{j,\mu}$ takes a different form depending on $j$ and $\mu$.

For $j$ even and $(j,\mu)\neq (N_t-1,t)$,
\begin{equation}
F_{j,\mu}=U_{j,\mu}
\left(-\frac{1}{4}(\Aoe X)^1_{j+\mu} (X^\dag)^0_j -
\left(\frac{1}{4} X^1_j ((\Aoe X)^\dag)^0_{j+\mu} \right)^\trasp \right) \ ;
\label{F1}
\end{equation}
for $j$ even and $(j,\mu)=(N_t-1,t)$,
\begin{equation}
F_{j,\mu}=U_{j,\mu}
\left(-\frac{1}{4}(\Aoe X)^0_{j+\mu} (X^\dag)^0_j +
\left(\frac{1}{4} X^1_j ((\Aoe X)^\dag)^1_{j+\mu} \right)^\trasp \right) \ ;
\end{equation}
for $j$ odd and $(j,\mu)\neq (N_t-1,t)$,
\begin{equation}
F_{j,\mu}=U_{j,\mu}
\left(\frac{1}{4} X^0_{j+\mu} ((\Aoe X)^\dag)^1_j +
\left(\frac{1}{4} (\Aoe X)^0_j (X^\dag)^1_{j+\mu}\right)^\trasp \right) \ ;
\end{equation}
for $j$ odd and $(j,\mu)=(N_t-1,t)$,
\begin{equation}
F_{j,\mu}=U_{j,\mu}
\left(\frac{1}{4} X^1_{j+\mu} ((\Aoe X)^\dag)^1_j -
\left(\frac{1}{4} (\Aoe X)^0_j (X^\dag)^0_{j+\mu}\right)^\trasp \right) \ .
\label{F4}
\end{equation}
Since we determined completely the elements of the matrices $\Aeo$ and $\Aoe$
in section~2.1, these expressions allow us to
obtain $F_{j,\mu}$ as a function of the links $U$ for every $j$ and $\mu$.

Coming back to Eq.~(\ref{Hamiltonian}), the condition $\dot{\cal H}=0$
gives
\begin{equation}
\dot{\cal H}=0=\sum_{j,\mu} \tr 
\left(\dot{H}_{j,\mu} H_{j,\mu} + 
\frac{\beta}{6}(i H_{j,\mu} U_{j,\mu} V_{j,\mu} + {\mathrm{H.c.}})-
(i H_{j,\mu} F_{j,\mu} + {\mathrm{H.c.}}) \right),
\end{equation}
where $V_{j,\mu}$ is the sum of staples, or products of the other three
matrices in the plaquettes containing $U_{j,\mu}$, and arises from the
differentiation of the Wilson action $S_g$. Of course, the staples
at the border of the lattice contain links defined by the boundary
conditions ($C^\star$ or periodic, depending on the direction).
The final solution for $\dot{H}_{j,\mu}$ is
\begin{equation}
i\dot{H}_{j,\mu}=\left[\frac{\beta}{3}U_{j,\mu}V_{j,\mu}-
2F_{j,\mu}\right]_{\mathrm TA}\ ,
\end{equation}
where the subscript TA indicates the traceless anti-Hermitian part
of the matrix:
\begin{equation}
Q_{\mathrm TA}=\frac{1}{2}(Q-Q^\dag)-\frac{1}{6}\tr(Q-Q^\dag) \ .
\end{equation}

\subsection{Reducing the number of flavours: the Hybrid algorithm}
\noindent
Because of Eq.~(\ref{pf}), the HMC algorithm that we have just described
simulates eight fermion flavours in the continuum. In order to come back
to four flavours, we note that
\begin{equation}
\mathrm{Pf}(A)=\pm (\det A)^{1/2}
\label{sign}
\end{equation}
(we saw in Eq.~(\ref{detApositive}) that $\det A$ is a positive number).

One can simulate $(\det A)^{1/2}$ by reverting to an approximate algorithm. 
A suitable
choice is the $R$ algorithm,\cite{gottlieb} in which 
discretization errors in the molecular dynamics part
are of ${\cal O}(\Delta \tau ^2)$.

On a lattice closed with $C^\star$-BC,
the full QCD action in the presence of $N_f$ families of degenerate continuum
fermions can be written as follows:
\begin{equation}
Z = \int ({\cal D} U)\, [\det (A^\dag A)_{\mathrm{e}}]^{N_f/8} \, e^{-S_g(U)} =
 \int ({\cal D} U)\, e^{-S_g(U) + \frac{N_f}{8} \tr 
\log (A^{\dag}A)_{\mathrm{e}}} \ ,
\end{equation}
where $(A^{\dag}A)_{\mathrm{e}}$ is the restriction of $A^{\dag}A$ to the 
even lattice sites.

Since the implementation of the $R$ algorithm in the present case reduces
to simple adaptation of a standard technique to the system described by the
equation of motion obtained in the previous subsection, we do not elaborate
further on this point. However, we still have the ``sign problem'' of
Eq.~(\ref{sign}). The sign of the Pfaffian can be included by 
reweighting the expectation values according to
\begin{equation}
\langle {\mathcal O}\rangle = 
\frac{\langle {\mathcal O} \cdot \mathrm{sign Pf}(A)\rangle_+}
{\langle \mathrm{sign Pf}(A)\rangle_+}\ ,
\end{equation}
where $\langle\cdots\rangle_+$ means that the expectation values have been
obtained simulating $+(\det A)^{1/2}$. 
One then needs to monitorize the sign of the Pfaffian depending on the
gauge configurations. Some techniques to do that are explained in Ref.~9.

\setcounter{equation}{0}
\setcounter{section}{4}
\section{Conclusions}
\noindent
$C^\star$ boundary conditions are interesting to study some spontaneous
symmetry breaking aspects of QCD. They are relevant when one analyses
confinement through monopole condensation. We have shown in this work how these
boundary conditions can be implemented to carry out a lattice simulation of
full QCD with staggered fermions. We have proved that the common even-odd
trick used to avoid the fermion redoubling produced by the introduction 
of the pseudofermionic field can be applied also to this case. However, 
there is an additional redoubling which forces to work with a 
minimum number of 
eight flavours with the usual Hybrid Monte Carlo algorithm, which we have
adapted to this case. An alternative to avoid that is to consider a non-exact
algorithm, which can be applied to any number of flavours. 

These algorithms have been implemented and are presently running on an
APE Quadrics machine to explore monopole condensation in full QCD.

\nonumsection{Acknowledgments}
\noindent
This work has been partially supported by EU TMR program
ERBFMRX-CT97-0122, Italian MURST and
PPARC Grant PPA/G/0/1998/00567. We thank valuable comments from
Isabel Campos, Simon Hands, Pilar Hern\'andez and Giampiero Paffuti.

\nonumsection{References}

\appendix

\noindent
We will show that $\det A$ is a real number, or, equivalently, 
that $\det A^\dag  = \det A $. This property is required in order
that one can use the matrix  $A^\dag A$ to introduce 
the pseudofermionic field.

Let us consider $A$ in the form
\begin{equation}
A=\left(\begin{array}{cc} B & -m -D^\trasp \\ 
m + D & B^\dag \end{array}\right),
\end{equation}
where  $D_{i,j} = {\tilde M}_{i,j} - m \delta_{i,j}$.

Using the general property of the determinant of a square
block matrix reported in Eq. (\ref{blockexp}) and exchanging\footnote{This 
corresponds to an even number of row exchanges, since $A$ has dimension
$2 N$ and $N$, being the dimension of the fermion matrix with periodic
boundary conditions, is always an even number. Therefore $\det A$ 
does not change under this operation.} \  the two rows of blocks of $A$ we 
can rewrite
\begin{eqnarray}
 \det A &=&  \det \left(\begin{array}{cc} m + D & B^\dag \\ 
B & -m -D^\trasp \end{array}\right) \nonumber \\
&=&  \det (m + D)
 \det (-m -D^\trasp -B (D + m)^{-1} B^\dag ) \ .
\label{deta}
\end{eqnarray}

Remembering that $D^\dag = - D$, we can write $A^\dag$ in the form
\begin{equation}
A^\dag=\left(\begin{array}{cc} B^\dag & m -D \\ 
-m + D^\trasp & B \end{array}\right),
\end{equation}
so that, exchanging the two columns of blocks in $A^\dag$ and using
again Eq. (\ref{blockexp}), we can write
\label{detadag}
\begin{eqnarray}
 \det A^\dag &=&  \det \left(\begin{array}{cc} m - D & B^\dag \\ 
B & -m  + D^\trasp \end{array}\right) \nonumber \\
&=&  \det (m - D)
 \det (-m +D^\trasp -B (-D + m)^{-1} B^\dag ) \ .
\label{deta2}
\end{eqnarray}

After extracting a factor $(-1)^N m^{2N} =  m^{2N}$ from both
$\det A$ and $\det A^\dag$ and defining $\alpha = 1/m$, we can write
\begin{eqnarray}
 \det A &=& m^{2N} 
 \det (1 + \alpha D) \det (1 + \alpha D^\trasp + \alpha^2 B (1 + \alpha D)^{-1} B^\dag ) \nonumber \\
 \det A^\dag &=& m^{2N} 
 \det (1 - \alpha D) \det (1 - \alpha D^\trasp + \alpha^2 B (1 - \alpha D)^{-1} B^\dag )
\label{detalpha}
\end{eqnarray}

It clearly appears from Eq. (\ref{detalpha}) that $\det A^\dag$ is obtained
from $\det A$ by changing the sign of $\alpha$. Therefore, in order to
show that $\det A = \det A^\dag$, it is sufficient to show that $\det A$
is an even function of $\alpha$.

Let us consider $\det  (1 + \alpha D)$ at first. We can expand
the determinant as follows:
\begin{eqnarray}
\det  (1 + \alpha D) &=& \exp \left( \tr\ln (1 + \alpha D) \right) \nonumber \\
&=& \exp \left( - \tr \sum_{k = 1}^\infty \frac{(-1)^k}{k} (\alpha D)^k \right).
\label{logexp}
\end{eqnarray}
It is easy to see that the trace of the product of an odd number of $D$
matrices is zero.
Indeed $D$ only connects nearest neighbour lattice sites, so it is 
not possible to connect a site to itself using the product of an odd 
number of $D$ matrices. Therefore only even powers of $\alpha$ appear
in the expansion in Eq. (\ref{logexp}) and  $\det  (1 + \alpha D)$
is an even function of $\alpha$.

Let us consider now $ \det (1 + \alpha D^\trasp + \alpha^2 B (1 + \alpha D)^{-1} B^\dag )$, which we rewrite as  $ \det (1 + P(\alpha)) $, where 
\begin{equation}
P(\alpha) =  \alpha D^\trasp + \alpha^2 B \left(   
\sum_{k = 0}^\infty (-1)^k \alpha^k D^k  \right)  B^\dag\; .
\end{equation}
We notice that the matrix $P(\alpha)$ is expressed as series
expansion in $\alpha$, where the coefficient of the $k$-th term
is a homogeneous polynomial of degree $k$ in the matrices 
$B,B^\dag,D$ and $D^\trasp$.
Therefore, expanding again the determinant as
\begin{eqnarray}
\det  (1 + P(\alpha)) &=& \exp \left( \tr\ln (1 + P(\alpha)) \right) 
\nonumber \\
&=& \exp \left( - \tr \sum_{k = 1}^\infty \frac{(-1)^k}{k} P (\alpha)^k \right),
\label{logexp2}
\end{eqnarray}
we see that  $ \det (1 + P(\alpha)) $ can be expanded as a power series in 
$\alpha$ and that the coefficient of the $k$-th term is the trace of a 
homogeneous polynomial of degree $k$ in the matrices $B,B^\dag,D$ and $D^\trasp$: since these matrices only connect nearest neighbour sites, the trace
is zero for $k$ odd. Therefore also in this case the determinant
is an even function of $\alpha$.

We conclude that $\det A$, being the product of even functions of $\alpha$,
is also an even function of $\alpha$, and therefore, from Eq. (\ref{detalpha}),
$\det A = \det A^\dag$.

\appendix

We will give here a proof, alternative to that presented in Ref.~8,
of the equality of the determinants of the submatrices on even
and odd sites of $M^\dag M$ in the case with standard boundary conditions.

In the standard case the fermion matrix $M$, defined in Eq. (\ref{fmatrix}),
has the following form
\begin{equation}
M=\left(\begin{array}{cc} m & \frac{1}{2} \Deo \\ 
\frac{1}{2} \Doe & m \end{array}\right)\; ,
\end{equation}
and it is easily shown that $\Deo^\dag = - \Doe$. Using this last
property it follows that
\begin{equation}
M^\dag M = \left(\begin{array}{cc} m^2 - \frac{1}{4} \Deo \Doe & 0\\ 
0 &  m^2 - \frac{1}{4} \Doe \Deo \end{array}\right).
\end{equation}
Using the decomposition of the determinant given in Eq. (\ref{blockexp}),
it is easy to show that
\begin{equation}
\det M = \det  \left( m^2 - \frac{1}{4} \Deo \Doe \right) \; .
\end{equation}
On the other hand, if we exchange even and odd  variables in $M$
before applying Eq. (\ref{blockexp}), we obtain
\begin{equation}
\det M = \det  \left( m^2 - \frac{1}{4} \Doe \Deo \right) \; ,
\end{equation}
and therefore
\begin{equation}
 \det  \left( m^2 - \frac{1}{4} \Doe \Deo \right) = 
 \det  \left( m^2 - \frac{1}{4} \Deo \Doe \right)\; .
\end{equation}

\end{document}